\documentclass[twocolumn,english,prd,superscriptaddress,nofootinbib,floatfix,showpacs]{revtex4-1}

\usepackage{float,graphicx}
\usepackage[latin1]{inputenc}
\usepackage[T1]{fontenc}
\usepackage{amsmath}
\usepackage{amssymb}
\usepackage{pstricks}
\usepackage{graphicx}
\usepackage{hyperref}
\usepackage{latexsym}
\usepackage{epsfig}
\usepackage{amssymb}
\usepackage[latin1]{inputenc}
\usepackage[T1]{fontenc}
\usepackage{amsmath}
\usepackage{pstricks}
\usepackage{graphicx}

\begin{document}

\title{The Cosmological Supersymmetron}

\author{Philippe Brax$^{\beta,}$}
\affiliation{Institut de Physique Theorique, CEA, IPhT, CNRS, URA 2306, F-91191Gif/Yvette Cedex, France}
\author{Anne-Christine Davis$^{\delta,}$}
\affiliation{DAMPT, Cambridge University, UK}
\author{Hans A. Winther$^{\omega,}$}
\affiliation{Institute of Theoretical Astrophysics, University of Oslo, 0315 Oslo, Norway}

\begin{abstract}
Recently, a supersymmetric model of dark energy coupled to cold dark matter, the supersymmetron, has been  proposed. In the absence of cold dark matter, the supersymmetron field converges to a supersymmetric minimum with a vanishing cosmological constant. When cold dark matter is present, the supersymmetron evolves to a matter dependent minimum where its energy density does not vanish and could lead to the  present acceleration of the Universe. The supersymmetron generates a short ranged fifth force which evades gravitational tests. It could lead to observable signatures on structure formation due to a very strong coupling to dark matter. We investigate the cosmological evolution of the field, focusing on the linear perturbations and the spherical collapse and find that observable modifications in structure formation can indeed exist. Unfortunately, we find that when the growth-rate of perturbations is in agreement with observations, an additional  cosmological constant is required to account for dark energy. In this case, effects on large scale structures are still present at the non-linear level which are investigated using the spherical collapse approach.
\end{abstract}


\maketitle

{\let\thefootnote\relax\footnotetext{
\\$^{\beta}$	Email address: \href{mailto:philippe.brax@cea.fr}{\nolinkurl{philippe.brax@cea.fr}}
\\$^{\delta}$	Email address: \href{mailto:A.C.Davis@damtp.cam.ac.uk}{\nolinkurl{a.c.davis@damtp.cam.ac.uk}}
\\$^{\omega}$	Email address: \href{mailto:h.a.winther@astro.uio.no}{\nolinkurl{h.a.winther@astro.uio.no}}}}

%%%%%%%%%%%%%%%%%%%%%%%%%%%%%%%%%%
%%%%%%%%%%%%%%%%%%%%%%%%%%%%%%%%%%
%%%%%%%%%%%%%%%%%%%%%%%%%%%%%%%%%%
%% INTRODUCTION
%%%%%%%%%%%%%%%%%%%%%%%%%%%%%%%%%%
%%%%%%%%%%%%%%%%%%%%%%%%%%%%%%%%%%
%%%%%%%%%%%%%%%%%%%%%%%%%%%%%%%%%%

\section{Introduction}
Dark energy, the component responsible for the late time acceleration of our Universe, is currently well described by a cosmological constant in the $\Lambda$CDM concordance model. $\Lambda$CDM has been very successful in explaining a large range of observations probing a vast range in length scales, but from a theoretical point of view the model suffers from  the fine-tuning problem and the coincidence problem \cite{Copeland:2006wr}. This has led  to  more general models for dark energy. Scalar field models have been particularly popular over the last decade, and are predicted to exist in many theories of high energy physics, like string theory and supergravity (see e.g. \cite{ss_book,Linde:2007fr} and references therein).

However, many of the dark energy models that have been constructed so far suffer from  problems akin to the ones they are trying to solve or introduce new issues themselves. At best they can be treated as low energy field theories valid  well below the electron mass, corresponding to the very late phase of the Universe. Hence these models need to be embedded in a better defined theory whose ultra violet behaviour is under control. So far, no such complete scenario has been constructed. Dark energy models also seem to require the existence of a very light scalar field whose coupling to matter leads to a long ranged fifth force whose presence is at odds with current gravitational tests. Screening mechanisms \cite{Vainshtein:1972sx,Deffayet:2001uk,ArkaniHamed:2002sp,Khoury:2003rn,Khoury:2003aq,Hinterbichler:2010es,Khoury:2010xi} have been invoked in order to alleviate this problem. Axion-like particles with derivative couplings to matter are also possible candidates \cite{Choi:1999xn}.

On the other hand, it could well be that the dark sector of the Universe, composed of the still undiscovered Cold Dark Matter (CDM) and Dark Energy (DE), could be  described by a globally supersymmetric theory \cite{Copeland:2000vh,Brax:1999gp,Brax:1999yv}. In such a case the vanishingly small amount of dark energy which is necessary to generate the acceleration of the Universe could result from a small cosmological breaking of supersymmetry due to the non-zero CDM energy density. Such a scenario would naturally lead to a close relationship between the dark energy and the CDM energy densities. Of course, one should also ensure that corrections to the globally supersymmetric scalar potential coming from the soft supersymmetry breaking in the MSSM sector do not spoil the CDM-DE correspondence and the properties of the scalar potential in the late time Universe.

Recently \cite{Brax:2011qs}, such a supersymmetric model of dark energy coupled to cold dark matter, the supersymmetron, was proposed by two of us. In the absence of cold dark matter, the supersymmetron converges to a supersymmetric minimum with a vanishing cosmological constant. When cold dark matter is present, the supersymmetron evolves to a matter dependent minimum where its energy density does not vanish and can contribute to the dark energy budget of the Universe.

The supersymmetron generates a short ranged fifth force between the CDM and the DE which evades gravitational tests, but could lead to observable signatures on structure formation as found in similar modified gravity theories \cite{Davis:2011pj}-\cite{Oyaizu:2008tb}.

In this paper we  analyse the cosmological evolution of the supersymmetron at the background level and the evolution of dark matter perturbations in the linear and non-linear regime. The non-linear regime is studied by using spherical collapse. Due to the highly non-linear behaviour of the field during the spherical collapse we are able to extract constraints on the model parameters, which are then used to constrain the background cosmology. The spherical collapse model has been previously used in models with a simple Yukawa-type modification of gravity, in the so-called $f(R)$ / chameleon models \cite{Schmidt:2008tn,Brax:2010tj,Brax:2005ew}, in brane-world cosmologies \cite{Schmidt:2009yj} and in models which allow for dark energy fluctuations \cite{Mota:2004pa}-\cite{Valageas:2010ki}. We find that a cosmological constant (CC) must be included in the model and that linear perturbations do not deviate from their $\Lambda$CDM counterparts. On the other hand, non-linear effects are significant on astrophysical scales.

The outline  of this paper is as follows. In Sec.~(\ref{sec1}) we present the supersymmetric formulation of the model, in Sec.~(\ref{sec2}) we derive static solutions to the field equation and then in Sec.~(\ref{sec3}) we study the cosmological evolution of the supersymmetron including the cosmological background evolution, linear perturbations and the  spherical collapse. In Sec.~(\ref{sec4}) we revisit the original mass scales of the model before summarizing and concluding in Sec.~(\ref{sec5}).

%%%%%%%%%%%%%%%%%%%%%%%%%%%%%%%%%%
%%%%%%%%%%%%%%%%%%%%%%%%%%%%%%%%%%
%%%%%%%%%%%%%%%%%%%%%%%%%%%%%%%%%%
%% SUPERSYMMETRON
%%%%%%%%%%%%%%%%%%%%%%%%%%%%%%%%%%
%%%%%%%%%%%%%%%%%%%%%%%%%%%%%%%%%%
%%%%%%%%%%%%%%%%%%%%%%%%%%%%%%%%%%

\section{The Supersymmetron}\label{sec1}
\subsection{Supersymmetric Formulation}
In globally supersymmetric models of the scalar sector, models are specified by their Kahler potential and the superpotential. With these two functions, we can construct the scalar potential and the kinetic term for the fields.
For the supersymmetron we have
\begin{align}
K(\phi,\overline{\phi},\phi_{\pm},\overline{\phi}_{\pm}) &= \left|\phi_+\right|^2 + \left|\phi_-\right|^2 +  \frac{\Lambda_1^2}{2}\left|\frac{\phi}{\Lambda_1}\right|^{2\beta}\\
W(\phi,\phi_{\pm}) &= m\left(1+\frac{g\phi}{m}\right)\phi_+\phi_- \\
&+ \frac{\Lambda_0^3\beta}{\sqrt{2}\alpha}\left(\frac{\phi}{\Lambda_0}\right)^{\alpha}+\frac{\Lambda_2^3}{\sqrt{2}}\left(\frac{\phi}{\Lambda_2}\right)^{\beta}
\end{align}
where $\phi$ is the dark energy superfield, $\phi_{\pm}$ is the CDM particles and $\Lambda_i$  are some (for now) unspecified mass-scales.

The kinetic term follows from
\begin{align}
\mathcal{L}_{\rm kin} &= K_{\phi\overline{\phi}}(\partial\phi)^2 = \frac{\kappa(\phi)^2}{2}(\partial\phi)^2\\
\kappa(\phi) &= \beta\left(\frac{\phi}{\Lambda_1}\right)^{\beta-1}
\end{align}
and the  scalar potential is given by the F-term
\begin{align}
V_F = K^{\phi\overline{\phi}}|\partial_{\phi}W|^2 = \left|\Lambda^2 + \frac{M^{2+n/2}}{\phi^{n/2}}\right|^2
\end{align}
where $n=2(\beta-\alpha)$ and the mass-scales $M$ and $\Lambda$ are given by
\begin{align}\label{scaleseq}
\Lambda^4 &= \left(\frac{\Lambda_1}{\Lambda_2}\right)^{2\beta-2}\Lambda_2^4\\
M^{n+4} &= \left(\frac{\Lambda_1}{\Lambda_0}\right)^{2\beta-2}\Lambda_0^{n+4}
\end{align}
Taking $\phi = |\phi|e^{i\theta}$, the scalar potential is seen to be minimized for $e^{\frac{in\theta}{2}}=-1$. The angular field $\theta$ is stabilised at this minimum with a mass which is always much greater than the gravitino mass \cite{Brax:2011qs} implying that
\begin{align}
V_F = \left(\Lambda^2 - \frac{M^{2+n/2}}{\phi^{n/2}}\right)^2
\end{align}
In the rest of this paper we will write $\phi$ instead of $|\phi|$ for simplicity. The potential is minimized, with vanishing potential energy, for $\phi = \phi_{\rm min}$ where
\begin{align}\label{phimineq}
\phi_{\rm min} = \left(\frac{M}{\Lambda}\right)^{\frac{4}{n}}M
\end{align}
Due to the coupling between $\phi$ and $\phi_{\pm}$ in the superpotential, the fermionic CDM particles  acquire a scalar field dependent mass
\begin{align}
m_f(\phi) = m\left(1+\frac{g\phi}{m}\right)
\end{align}
When the fermionic-CDM develops a non-vanishing number density $n_{\rm CDM} = \left<\psi_+\psi_-\right>$ in the early Universe we get a new contribution to the scalar potential
\begin{align}
V_{\rm eff} = V_F + \frac{g\phi}{m}\rho_{\rm CDM},~~~~~\rho_{\rm CDM} = mn_{\rm CDM}
\end{align}
which lifts the supersymmetric minimum and produces a non-zero dark energy component which can lead to the acceleration of our Universe.

%%%%%%%%%%%%%%%%%%%%%%%%%%%%%%%%%%
%% EFFECTIVE 4D MODEL
%%%%%%%%%%%%%%%%%%%%%%%%%%%%%%%%%%

\subsection{Effective $4$D Model}
The effective theory for the sypersymmetron can be viewed as a scalar tensor theory where CDM particles follow geodesics of the rescaled
metric $\tilde{g}_{\mu\nu} = g_{\mu\nu}A(\phi)$ where
\begin{align}
A(\phi) = 1+\frac{g\phi}{m}
\end{align}
The effective $4$D action describing the dynamics of the supersymmetron is given by
\begin{align}
S_{\rm eff} &= \int d^4x\sqrt{-g}\left[ \frac{R}{2}M_{\rm pl}^2 - \frac{\kappa(\phi)^2}{2}(\partial\phi)^2 - V_F(\phi) \right]\nonumber\\
&+ S_{\rm CDM}(A^2(\phi)g_{\mu\nu};\psi)
\end{align}
where $g$ is the determinant of the metric $g_{\mu\nu}$, $M_{\rm pl}^2 \equiv \frac{1}{8\pi G}$ is the reduced Planck mass and $S_{\rm CDM}$ is the dark matter action. If a coupling to baryons is included, the large mass of the supersymmetron field will ensure that this field would be practically invisible in local experiments.

%%%%%%%%%%%%%%%%%%%%%%%%%%%%%%%%%%
%% REPARAMETRIZATION OF MODEL PARAMETERS
%%%%%%%%%%%%%%%%%%%%%%%%%%%%%%%%%%

\subsection{Reparametriziation of the model parameters}
In this subsection we rewrite the original mass-scales of the model in terms of some more intuitive physical quantities which will simplify our analysis.
\\\\
The coupling of the supersymmetron to dark matter $A(\phi)$ can be written
\begin{align}\label{conformal_coupling}
A(\phi) = 1 + x \left(\frac{\phi}{\phi_{\rm min}}\right)
\end{align}
where
\begin{align}\label{Xeq}
x \equiv \frac{g\phi_{\rm min}}{m}
\end{align}
is a dimensionless parameter which parameterises the coupling strength of the supersymmetron to matter.
\\\\
We further introduce the density
\begin{align}\label{Lambdaeq}
\rho_{\infty} \equiv \frac{n \Lambda^4}{x}
\end{align}
and 
\begin{align}
\rho_{\infty}\equiv \rho_{CDM}^0(1+z_{\infty})^3
\end{align}
which is the CDM density (and the corresponding redshift) when the field $\phi$ reaches the vicinity of the supersymmetric minimum $\phi_{\rm min}$.
\\\\
The mass of the supersymmetron after having converged to the supersymmetric minimum is given by
\begin{align}
m_{\infty}^2 = \frac{\rho_{\infty} x n}{2\beta^2\varphi_{\rm min}^2}
\end{align}
Constraints from supersymmetry breaking require $m_{\infty} \gg m_{3/2}$ where $m_{3/2}$ is the gravitino mass and is typically much larger than $1$eV \cite{Nilles:1983ge}. We will therefore require
\begin{align}\label{mass_constraint}
m_{\infty} \gg {\cal O} (1) {\rm eV}
\end{align}
which is our first constraint. The three physical parameters $\{\rho_{\infty}, x, m_{\infty}\}$ together with the two indices $\{n,\beta\}$  completely characterise the effective model. In the end we will go back and compare our results with the original mass-scales.

When studying the cosmological dynamics of the model, it is convenient to introduce the canonically normalized field $\varphi$ via
\begin{align}
d\varphi = \kappa(\phi)d\phi~~~~\to~~~~\varphi(\phi) = \Lambda_1\left(\frac{\phi}{\Lambda_1}\right)^{\beta}
\end{align}
In terms of $\varphi$ the potential and coupling becomes
\begin{align}\label{potential}
V_F(\varphi) &= \Lambda^4\left(1 - \left(\frac{\varphi_{\rm min}}{\varphi}\right)^{\frac{n}{2\beta}}\right)^2\\
A(\varphi) &= 1 + x \left(\frac{\varphi}{\varphi_{\rm min}}\right)^{\frac{1}{\beta}}
\end{align}
where $\varphi_{\rm min} = \varphi(\phi_{\rm min})$.

%%%%%%%%%%%%%%%%%%%%%%%%%%%%%%%%%%
%% SUPERSYMMETRON DYNAMICS
%%%%%%%%%%%%%%%%%%%%%%%%%%%%%%%%%%

\subsection{Supersymmetron dynamics}
The field equation for $\varphi$ follows from a variation of Eq.~(\ref{fieldequation}) with respect to $\varphi$ and reads
\begin{align}\label{fieldequation}
\square\varphi = V_{\rm eff, \varphi}
\end{align}
where the effective potential is given by
\begin{align}\label{veff}
V_{\rm eff}(\varphi) = V_F(\varphi) + (A(\varphi)-1)\rho_{\rm CDM}
\end{align}
The minimum $\varphi_{\rho}$ of the effective potential is determined by
\begin{align}
\left(\frac{\varphi_{\rm min}}{\varphi_{\rho}}\right)^{\frac{n+1}{\beta}}-\left(\frac{\varphi_{\rm min}}{\varphi_{\rho}}\right)^{\frac{n+1}{2\beta}} = \frac{\rho_{\rm CDM}}{\rho_{\infty}}
\end{align}
and has the approximate solution
\begin{align}\label{minimum}
\frac{\varphi_{\rho}}{\varphi_{\rm min}} \simeq \left\{
	\begin{array}{ll} 1 & ~~~~\rho_{\rm CDM} \lesssim \rho_{\infty}\\
		\left(\frac{\rho_{\infty}}{\rho_{\rm CDM}}\right)^{\frac{\beta}{n+1}} & ~~~~\rho_{\rm CDM} \gg \rho_{\infty}
	\end{array}\right.
\end{align}
A non-zero dark matter condensate is seen to lower the minimum from the supersymmetric minimum. The energy density associated with the supersymmetron is
\begin{align}
V_{\rm eff}(\varphi_{\rho}) \simeq \left\{
	\begin{array}{ll}
		\frac{x\rho_{\rm CDM}(1+n)}{n} \left(\frac{\rho_{\infty}}{\rho_{\rm CDM}}\right)^{\frac{1}{n+1}} & ~~~~\rho_{\rm CDM} \gg \rho_{\infty}\\
		x\rho_{\rm CDM} &  ~~~~\rho_{\rm CDM} \lesssim \rho_{\infty}
	\end{array}\right.
\end{align}
which for small $n$ and $\rho_{\rm CDM} \gg \rho_{\infty}$ behaves like a cosmological constant, but evolves as CDM after the field has converged to the supersymmetric minimum. This means that if the supersymmetron accounted for all dark energy,  then acceleration would be a transient phenomenon.

The mass of the field, $m_{\varphi}^2 \equiv V_{\rm eff,\varphi\varphi}$, is given by
\begin{align}
m_{\varphi}^2 = m_{\infty}^2&\left[ \frac{2(n+\beta)}{n}\left(\frac{\varphi_{\rm min}}{\varphi_{\rho}}\right)^{\frac{n}{\beta}+2} - \frac{(n+2\beta)}{n}\left(\frac{\varphi_{\rm min}}{\varphi_{\rho}}\right)^{\frac{n}{2\beta}+2} \right.\nonumber\\
&\left.+~~\frac{2(1-\beta)}{n}\frac{\rho_{\rm CDM}}{\rho_{\infty}}\left(\frac{\varphi_{\rm min}}{\varphi_{\rho}}\right)^{2-\frac{1}{\beta}} \right]
\end{align}
When the field follows the minimum of the effective potential this expression simplifies to
\begin{align}\label{mass_minimum}
\frac{m_{\rho}^2}{m_{\infty}^2} \simeq \left\{
	\begin{array}{ll}
		1 & ~~~~\rho_{\rm CDM} \lesssim \rho_{\infty}\\
		\frac{2(n+1)}{n}\left(\frac{\rho_{\rm CDM}}{\rho_{\infty}}\right)^{\frac{n+2\beta}{n+1}} & ~~~~\rho_{\rm CDM} \gg \rho_{\infty}
	\end{array}\right.
\end{align}
We see that the mass is always greater than the value at the supersymmetric minimum and from the constraint Eq.~(\ref{mass_constraint}) the mass is therefore always  greater than a few  eV's .

The conformal coupling Eq.~(\ref{conformal_coupling}) leads to a fifth-force (see e.g. \cite{Mota:2006fz,Brax:2010kv}) between dark matter particles, which in the non-relativistic limit (and per unit mass) is given by
\begin{align}\label{fifthforce}
\vec{F}_{\varphi} = \frac{d\log A(\varphi)}{d\varphi}\vec{\nabla}\varphi
\end{align}
This fifth-force will have an impact on structure formation which is investigated in the following sections.

%%%%%%%%%%%%%%%%%%%%%%%%%%%%%%%%%%
%%%%%%%%%%%%%%%%%%%%%%%%%%%%%%%%%%
%%%%%%%%%%%%%%%%%%%%%%%%%%%%%%%%%%
%% Static configurations
%%%%%%%%%%%%%%%%%%%%%%%%%%%%%%%%%%
%%%%%%%%%%%%%%%%%%%%%%%%%%%%%%%%%%
%%%%%%%%%%%%%%%%%%%%%%%%%%%%%%%%%%

\section{Static configurations}\label{static}\label{sec2}
In this section we derive the static spherical symmetric solutions for the supersymmetron, which we then use later on when studying the spherical collapse.

In a static spherical symmetric metric with weak gravitational fields the field equation Eq.~(\ref{fieldequation}) reads
\begin{align}
\frac{d^2\varphi}{dr^2}+\frac{2}{r}\frac{d\varphi}{dr} = V_{F,\varphi} + A_{,\varphi}\rho
\end{align}
We consider a spherical body of dark matter (a halo) with radius $R$ and a top-hat density profile
\begin{align}
\rho = \left\{
	\begin{array}{ll}
		\rho_c & r < R \\ \rho_b & r > R
	\end{array}\right.
\end{align}
and impose the standard boundary conditions
\begin{align}
&\frac{d\varphi(r\to0)}{dr} = \frac{d\varphi(r\to \infty )}{dr} = 0\\
&\varphi(r\to\infty) = \varphi_b = \varphi_{\rho}(\rho_b)
\end{align}
The mass at the minimum inside (outside) the body is denoted by $m_c$ ($m_b$).

Outside the halo we can linearize the field equation around the background value $\varphi_b$ with the solution
\begin{align}
\varphi(r) = \varphi_b - \frac{BR}{r}e^{-m_b(r-R)}~~~~r>R
\end{align}
and where the constant $B$ is determined by matching to the interior solution. The interior solutions are calculated below for several different cases.

%%%%%%%%%%%%%%%%%%%%%%%%%%%%%%%%%%
%% POINT PARTICLE SOLUTIONS
%%%%%%%%%%%%%%%%%%%%%%%%%%%%%%%%%%

\subsection{Point particle solutions}
We first look at the point particle solution, which can be found by first deriving the solution for fixed $R$ and then taking the limit $R\to 0$ with $M = \frac{4\pi}{3}R^3\rho_c$ fixed.

In this limit we expect the solution inside the body to be a very small perturbation of the background solution and we can assume $m_b R \ll 1$. A second order Taylor expansion in $r$ gives us the solution
\begin{align}
\varphi = \varphi(0) + \frac{A_{,\varphi_b}\rho_c r^2}{6}~~~~~r<R
\end{align}
Matching to the exterior solution at $r=R$ gives
\begin{align}
\varphi(0)  &= \varphi_b - \frac{A_{,\varphi_b}\rho_c R^2}{2}\\
BR &=  \frac{A_{,\varphi_b}M}{4\pi}
\end{align}
Taking the limit $R\to 0$ and using Eq.~(\ref{fifthforce}) we find that the gravity plus fifth-force potential is given by
\begin{align}\label{force_pot}
V(r) = \frac{GM}{r}\left(1+2(A_{,\varphi_b}M_{\rm pl})^2e^{-m_{\varphi}r}\right)
\end{align}
which is the same as the prediction from linear perturbation theory\footnote{By taking the Fourier transform of Eq.~(\ref{linear_pert}) we recover Eq.~(\ref{force_pot}), see e.g. \cite{amendola_shinji}.} as we will se later on. Contrary  to chameleons where this type of solution holds at linear scales, here the large mass of the supersymmetron means that this solution only applies for microscopic bodies.

%%%%%%%%%%%%%%%%%%%%%%%%%%%%%%%%%%
%% SMALL OVERDENSITY
%%%%%%%%%%%%%%%%%%%%%%%%%%%%%%%%%%

\subsection{Small overdensity}
Now we turn to the case where the size of the overdensity has to be taken into account. Note that we cannot make the approximation $m_b R \ll 1$ as the mass of the supersymmetron is generally very large
\begin{align}
\frac{1}{m_b} < \frac{1}{m_{\infty}} \ll \frac{1}{\text{eV}} \sim 10^{-6}m
\end{align}
Since we are interested in astrophysical sized overdensities, $R = \mathcal{O}(\text{Mpc})$, we will always have $m_b R \gg 1$.
\\\\
We take $\varphi = \varphi_0 + \delta\varphi$ and Taylor expand the field equation inside the body around $\varphi_0 \equiv \varphi(r=0)$:
\begin{align}
\frac{d^2\delta\varphi}{dr^2} + \frac{2}{r}\frac{d\delta\varphi}{dr} = V^{\rm eff}_{,\varphi_0} +m_0^2\delta\varphi,~~~~~  m_0^2 \equiv V^{\rm eff}_{,\varphi\varphi_0}
\end{align}
which gives the solution
\begin{align}
\varphi = \varphi_0 + \frac{ V^{\rm eff}_{,\varphi_0}}{m_0^2}\left(\frac{\sinh(m_0 r)}{m_0 r} -1\right)~~~~~r < R
\end{align}
Matching at $r=R$ and using $m_0R,m_b R \gg 1$ to simplify the analysis, we find
\begin{align}
\varphi_0 &\simeq \varphi_b  - \frac{V^{\rm eff}_{,\varphi_0}}{m_0^2}\left(\frac{\sinh(m_0 R)}{m_0 R} + \frac{\cosh(m_0 R)}{m_b R} \right)\\
B &\simeq \frac{\varphi_b-\varphi_0}{1+\frac{m_b}{m_c}} \simeq  \frac{\varphi_b-\varphi_0}{2}
\end{align}
We assume that $\varphi_0$ is just a small perturbation in the background: $\varphi_0 = \varphi_b - \delta\varphi$, and expand the above expression to first order in $\delta\varphi$. This leads to
\begin{align}
\delta\varphi \simeq \frac{(\rho_c-\rho_b)A_{,\varphi_b}}{m_b^2}
\end{align}
and
\begin{align}
B = \frac{(\rho_c-\rho_b)A_{,\varphi_b}}{2m_b^2}
\end{align}
Which gives a total force, $F = \frac{G_{\rm eff}\Delta M}{R^2}$ on a shell close to the edge of the  overdensity where
\begin{align}\label{geff_smalldelta}
\left(\frac{G_{\rm eff}}{G}\right) = 1 + \frac{6(A_{,\varphi_b}M_{\rm pl})^2}{(m_b R)}
\end{align}
which is suppressed compared to the point-particle solution. This solution is only valid when
\begin{align}
\delta\varphi \simeq \frac{(\rho_c-\rho_b)A_{,\varphi_b}}{m_b^2} \ll \varphi_b
\end{align}
Putting $\rho_c = (1+\Delta)\rho_b$ and using $\frac{\rho_b A_{,\varphi_b}}{\varphi_b} \sim m_b^2$ we see that this condition reduces to
\begin{align}
\Delta \ll 1
\end{align}
i.e. a very small overdensity.

%%%%%%%%%%%%%%%%%%%%%%%%%%%%%%%%%%
%% LARGE OVERDENSITY
%%%%%%%%%%%%%%%%%%%%%%%%%%%%%%%%%%

\subsection{Large overdensity}
For large overdensities we expect screening and we therefore look for  chameleon-like solutions \cite{Khoury:2003aq,Khoury:2003rn,Brax:2004qh,Brax:2004px,Mota:2006fz,Mota:2008fc,Tamaki:2008mf}. That is we assume that the field is very close to the minimum, $\varphi_c = \varphi_{\rho}(\rho_c)$, of the effective potential inside the body and the only variations of the field are in a thin-shell close to the surface. Linearizing the field equation about $\varphi_c$ leads to the solution
\begin{align}
\varphi(r) = \varphi_c + C\frac{\sinh(m_c r)}{m_c r}~~~~r<R
\end{align}
Due to the form of the field-equation for general $\beta \not = 1$ we cannot solve the equation in the thin-shell, but we will assume that this solution is valid all the way to $r=R$. This will be the case if the shell is very thin as found in chameleon theories \cite{Brax:2010kv,Tamaki:2008mf,Mota:2008fc}, and as we will see below this is the case for the supersymmetron when the density contrast of our overdensity is large.  In fact, we find that the supersymmetron is very similar in behaviour to strongly coupled chameleons as studied in \cite{Mota:2008fc}.

Matching the two solutions at $r=R$ we obtain
\begin{align}
B \simeq (\varphi_b- \varphi_c)
\end{align}
i.e. the solution found is the critical solution where the field almost does not change inside the body. We can rewrite this equation in the standard chameleon form by introducing the equivalent thin-shell factor
\begin{align}
\frac{\Delta R}{R} \equiv \frac{(\varphi_b- \varphi_c)}{6\beta_{\varphi_c}\Phi_c M_{\rm pl}^2},~~~~~\beta_{\varphi_c} = A_{,\varphi_c}M_{\rm pl}
\end{align}
where $\Phi_c = \frac{G\Delta M}{R} = \frac{(\rho_c-\rho_b)R^2}{6M_{\rm pl}^2}$ is the Newtonian potential of the overdensity.  The total force on a spherical shell close to the surface is now $F = \frac{G_{\rm eff}\Delta M}{R^2}$ where
\begin{align}\label{geff_sph}
\left(\frac{G_{\rm eff}}{G}\right) = 1+2\beta_{\varphi_c}^2 \left(\frac{3\Delta R}{R}\right)\times (1+m_b R)
\end{align}
This solution is valid as long as the quadratic term in the Taylor expansion of $V_{\rm eff,\varphi}$ around $\varphi_c$ is suppressed compared to the linear term at $r=R$. This condition turns into
\begin{align}
\frac{m_b}{m_c}\left(\frac{\varphi_b}{\varphi_c}\right) \ll 1~~~\to~~~\Delta \gg 1
\end{align}
We have not found an explicit solution for $\Delta \sim 1$, but if we take $\Delta\to 0$ in Eq.~(\ref{geff_sph}) we recover Eq.~(\ref{geff_smalldelta}). Thus the two approximation agree for $\Delta \sim 1$ and we will therefore use equation Eq.~(\ref{geff_sph}) as an approximation for the fifth-force for all $\Delta$.

For a body of fixed size $R$ the effective gravitational constant is seen to decrease with increasing $\rho_c$ demonstrating the chameleon-like behaviour and thus we recover the Newtonian regime for virialised halos, see Fig.~(\ref{geff_fig}).

%%%%%%%%%%%%%%%%%%%%%%%%%%%%%%%%%%
%%%%%%%%%%%%%%%%%%%%%%%%%%%%%%%%%%
%%%%%%%%%%%%%%%%%%%%%%%%%%%%%%%%%%
%% COSMOLOGICAL SUPERSYMMETRON
%%%%%%%%%%%%%%%%%%%%%%%%%%%%%%%%%%
%%%%%%%%%%%%%%%%%%%%%%%%%%%%%%%%%%
%%%%%%%%%%%%%%%%%%%%%%%%%%%%%%%%%%

\section{Cosmological Supersymmetron}\label{sec3}
In this section we discuss the cosmological evolution of the supersymmetron at the  background level, the linear perturbations and the spherical collapse.

%%%%%%%%%%%%%%%%%%%%%%%%%%%%%%%%%%
%% BACKGROUND COSMOLOGY
%%%%%%%%%%%%%%%%%%%%%%%%%%%%%%%%%%

\subsection{Background Cosmology}
The background evolution of the supersymmetron in a flat Friedmann-Lemaitre-Robertson-Walker (FLRW) metric
\begin{align}\label{frlwmetric}
ds^2 = -dt^2 + a(t)^2(dx^2+dy^2+dz^2)
\end{align}
is determined by the Friedman equation which in the late Universe reads
\begin{align}
3M_{\rm pl}^2H^2 = \rho_b + \rho_{\rm CDM} + \rho_{DE}
\end{align}
where $\rho_b$ is the baryon density, $\rho_{\rm CDM}$ the dark matter densiy and $\rho_{DE}$ is the dark energy density. In the following we will ignore the baryons and treat all matter as CDM. The CDM energy density is conserved implying that
\begin{align}
\dot{\rho}_{CDM} + 3H\rho_{\rm CDM} = 0
\end{align}
The DE density is given by the sum of the energy density in the supersymmetron and a cosmological constant (CC)
\begin{align}
\rho_{DE} = \rho_{\rm CC} +  \rho_{\varphi}
\end{align}
where $\rho_{\varphi} = \frac{\dot{\varphi}^2}{2} + V_F + (A(\varphi)-1)\rho_{\rm CDM}$.  We will later see that a non-zero CC is required to have a viable cosmology. This CC may come from supersymmetry breaking \cite{Brax:2011qs}.

\begin{figure}
\centering
\includegraphics[width=\columnwidth]{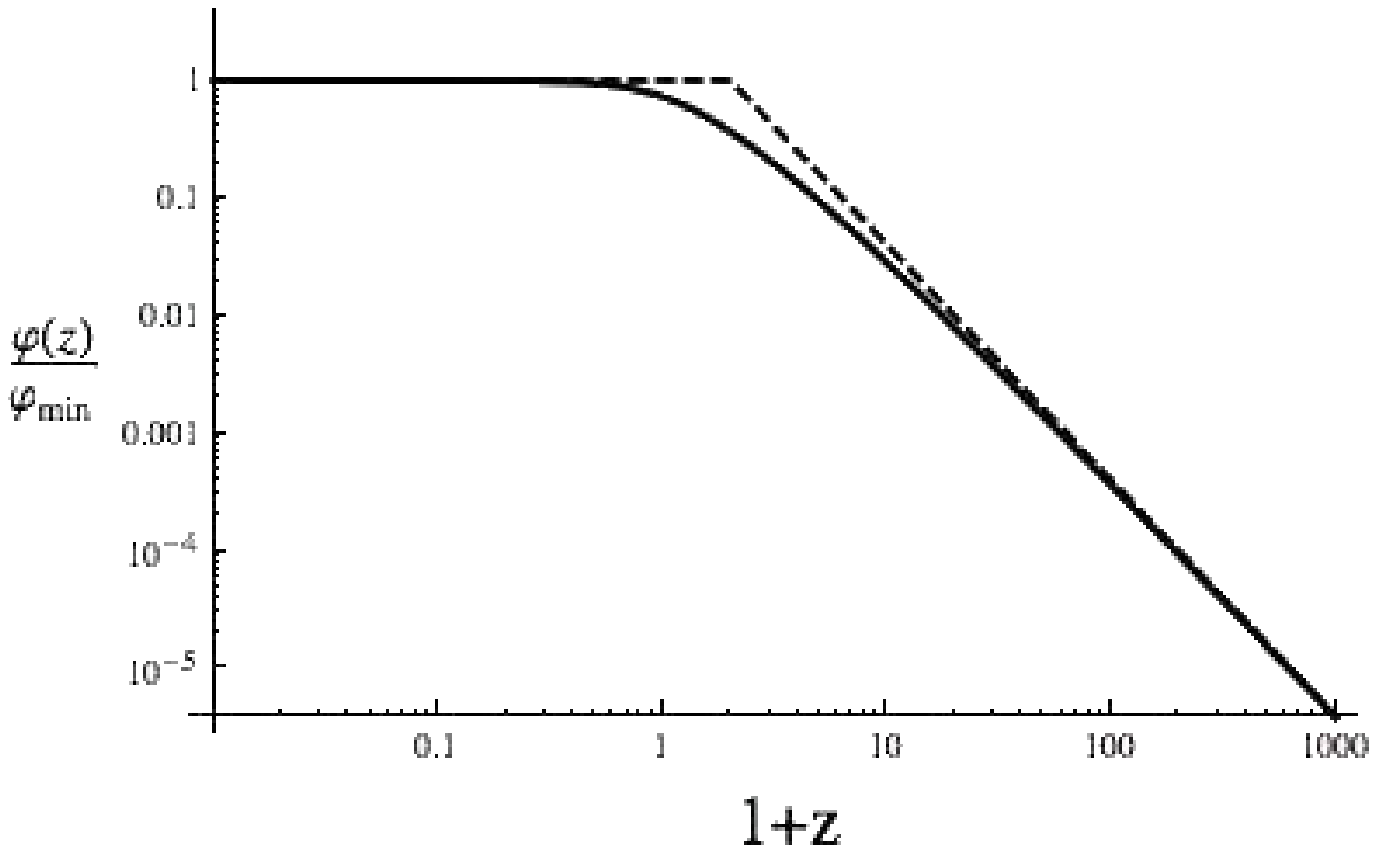}\\
\includegraphics[width=\columnwidth]{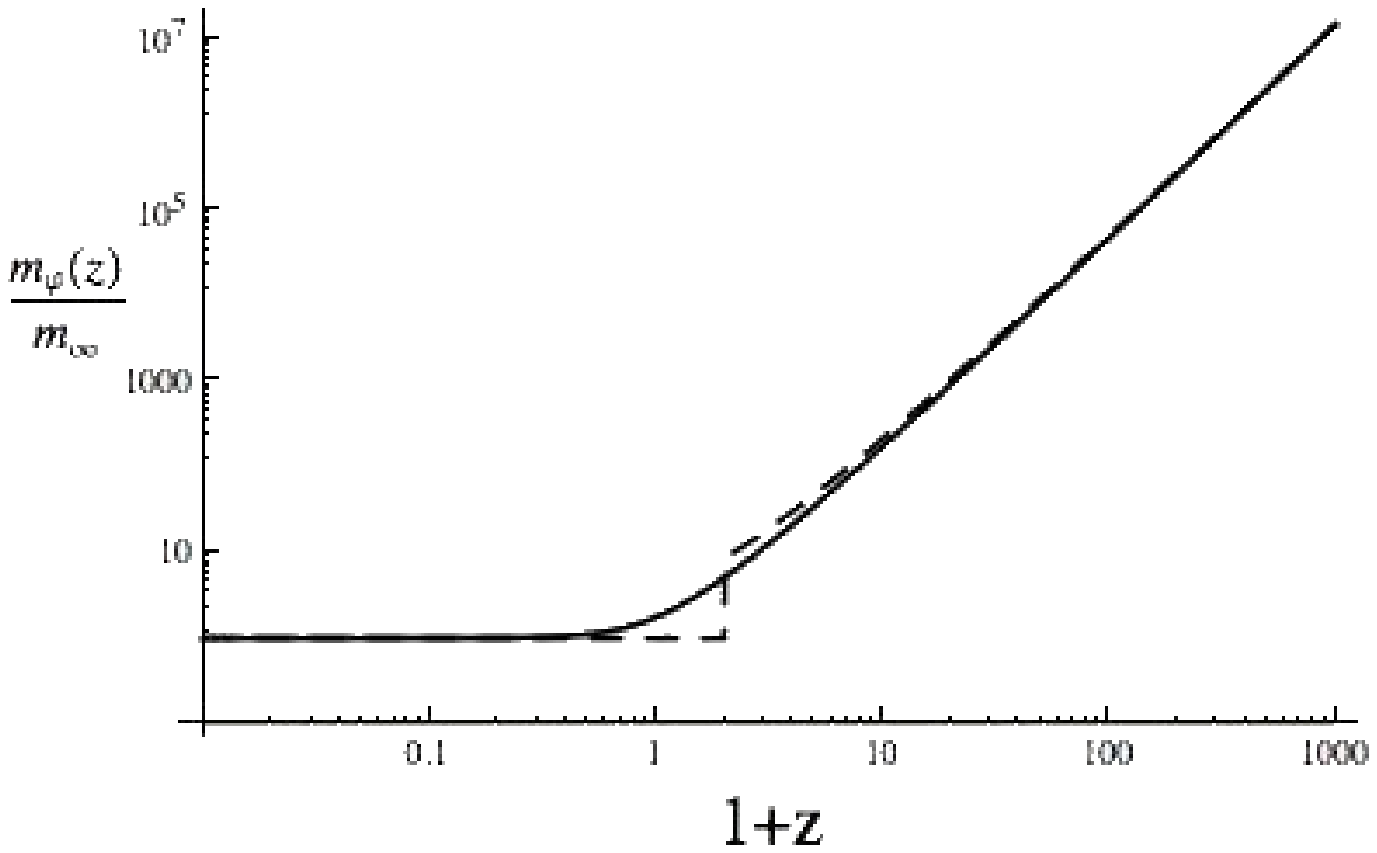}
\caption{The cosmological evolution of $\varphi_{\rho}$ (above) and $m_{\varphi}$ (below) as function of redshift. The dashed line shows the analytical approximation Eq.~(\ref{minimum}) (above) and Eq.~(\ref{mass_minimum}) (below). The supersymmetron parameters are $z_{\infty} = 1.0$, $\beta=1$ and  $n=0.5$. $x$ and $m_{\infty}$ does not have any influence on the evolution of the minimum and does therefore not need to be specified here.}
\label{phiminplot}
\end{figure}

The field equation Eq.~(\ref{fieldequation}) in the FLRW metric Eq.~(\ref{frlwmetric}) becomes
\begin{align}
\ddot{\varphi} + 3H\dot{\varphi} + V_{\rm eff,\varphi} = 0
\end{align}
The mass of the field is constrained by Eq.~(\ref{mass_constraint}) which means that $m_{\varphi} \gg H$ in the late Universe. The minimum $\varphi_{\rho}$ is therefore an attractor which the field follows. Along this attractor the kinetic term is negligible as $\frac{\dot{\varphi}^2}{2V_{\rm eff}} \sim \left(\frac{H}{m_{\varphi}}\right)^2 \ll 1$. In Fig.~(\ref{phiminplot}) we show the cosmological evolution of $\varphi_{\rho}$ and $m_{\varphi}$ with redshift.

\begin{figure}
\centering
\includegraphics[width=\columnwidth]{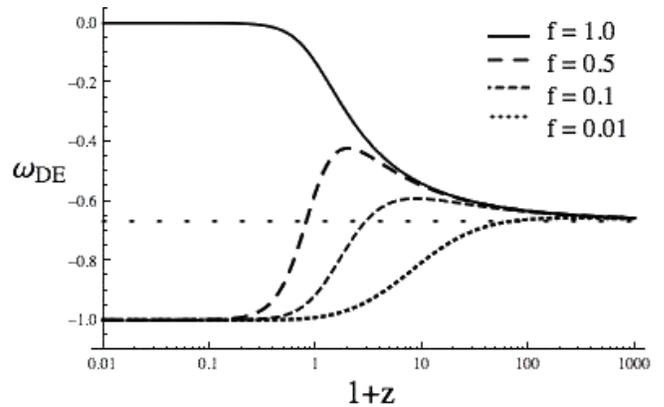}
\caption{The equation of state for the supersymmetron as function of redshift for four different values of $f = \frac{\Omega_{\varphi}}{\Omega_{\varphi}+\Omega_{\rm CC}}$: the fraction of dark energy in the supersymmetron to the total dark energy density today. The horizontal dotted line shows the analytical approximation $\omega_{\varphi} = -\frac{1}{n+1}$. The supersymmetron parameters are $z_{\infty} = 1.0$, $\beta=1$, $n=0.5$ and $x$ is fixed to give the desired $f$ in each case.}
\label{eqstate}
\end{figure}

When the supersymmetron follows the attractor we have
\begin{align}\label{rhomin}
\rho_{\varphi} \simeq \left\{
	\begin{array}{ll}
		x\rho_{\rm CDM}\frac{(n+1)}{n}\left(\frac{\rho_{\infty}}{\rho_{\rm CDM}}\right)^{\frac{1}{n+1}} & ~~~~\rho_{\rm CDM} \gg \rho_{\infty} \\
		x\rho_{\rm CDM} &~~~~ \rho_{\rm CDM} \lesssim \rho_{\infty}
	\end{array}\right.
\end{align}
The equation of state along the attractor is given by
\begin{align}
\omega_{\varphi} = \frac{p_{\varphi}}{\rho_{\varphi}} \simeq -\frac{V_F}{V_{\rm eff}} \simeq \left\{
	\begin{array}{ll}
	 	-\frac{1}{n+1} & ~~~~\rho_{\rm CDM} \gg \rho_{\infty}\\
	 	0 & ~~~~\rho_{\rm CDM} \ll  \rho_{\infty}
	 \end{array}\right.
\end{align}
To have acceleration of the Universe without a CC we need to impose $z_{\infty} > 0$ and $n < 2$. When the field converges to the supersymmetric minimum $p_{\varphi} \simeq - V_F \to 0$ and the acceleration stops. To have agreement with observations we need a non-zero CC, as was pointed out in \cite{Brax:2011qs}. With the inclusion of a CC, the dark energy equation of state is modified
\begin{align}\label{eom_de}
\omega_{\rm DE} = \frac{p_{\varphi}-\rho_{\rm CC}}{\rho_{\varphi}+\rho_{\rm CC}} \simeq \left\{
	\begin{array}{lr}
		\omega_{\varphi}& ~~~~\rho_{\varphi} \gg \rho_{\rm CC}\\
		-1 & ~~~~\rho_{\varphi} \ll  \rho_{\rm CC}
	\end{array}\right.
\end{align}
In Fig.~(\ref{eqstate}) we show the dark energy equation of state as function of redshift and $f \equiv \frac{\Omega_{\varphi}}{\Omega_{\varphi}+\Omega_{\rm CC}}$: the fraction of dark energy in the supersymmetron to the total dark energy density today.

To find out how large a contribution the supersymmetron we can have in the energy budget of our Universe we will first look at the linear perturbations to get a constraint on the model parameters and then apply these constraints to the background cosmology.

%%%%%%%%%%%%%%%%%%%%%%%%%%%%%%%%%%
%% LINEAR PERTURBATIONS
%%%%%%%%%%%%%%%%%%%%%%%%%%%%%%%%%%

\subsection{Linear perturbations}

The coupling of the supersymmetron Eq.~(\ref{conformal_coupling}) to dark matter leads to a fifth-force which will influence the growth of the linear perturbations and structure formation in general. The similarity of the model with chameleons yields that in high density regions the fifth-force will be screened as shown in Sec.~(\ref{static}).
\\\\
The growth of the dark matter perturbations $\delta = \frac{\delta\rho_{\rm CDM}}{\rho_{\rm CDM}}$ for sub-horizon scales are determined by (see e.g. \cite{Amendola:2003wa,Mota:2010uy,Brax:2004qh,Gannouji:2010fc})

\begin{align}\label{linear_pert}
\ddot{\delta} + 2H\dot{\delta} = \frac{3}{2}H^2\Omega_{\rm CDM}(a) \delta \left(\frac{G_{\rm eff}(k)}{G}\right)_{\rm lin}
\end{align}
where the effective gravitational constant is given by
\begin{align}
 \left(\frac{G_{\rm eff}(k)}{G}\right)_{\rm lin} &= 1 + \frac{2((\log A)_{,\varphi}M_{\rm pl})^2}{1 + \frac{a^2m_{\varphi}^2}{k^2}}\nonumber\\
 &\approx 1 +  \frac{2(A_{,\varphi}M_{\rm pl})^2}{m_{\varphi}^2}\frac{k^2}{a^2}
\end{align}
where the last equality comes from the fact that the supersymmetron is very heavy compared to astrophysical scales and where we have assumed $A(\varphi)-1 \ll 1$ (see Eq.~(\ref{a_constraint})).

In order to have signatures on the linear perturbations we need the coupling strength to satisfy $2(A_{,\varphi}M_{\rm pl})^2 \gg 1$, i.e. the supersymmetron must be very strongly coupled. It has been argued \cite{Bean:2007ny,Bean:2007nx} that an adiabatic instability exists in the regime, a point we will return to when discussing the non-linear evolution in the next section.

At early times, $\rho_{\rm CDM} \gg \rho_{\infty}$, we find
\begin{align}\label{geff_lin1}
 \left(\frac{G_{\rm eff}(k)}{G}\right)_{\rm lin} &\approx 1+\left(\frac{6}{(n+1)\Omega_{\rm CDM}^0}\right)\times x10^4 \nonumber\\
 & \times \left(\frac{k/a}{0.1h\text{Mpc}^{-1}}\right)^2 \times \left(\frac{1+z_{\infty}}{1+z}\right)^{\frac{3}{n+1}}
\end{align}
and as the field converges to the supersymmetric minimum we obtain
\begin{align}\label{geff_lin2}
 \left(\frac{G_{\rm eff}(k)}{G}\right)_{\rm lin} &\approx 1+\left(\frac{12}{n\Omega_{\rm CDM}^0}\right)\times \frac{x10^4}{(1+z_{\infty})^3} \nonumber\\
 & \times \left(\frac{k/a}{0.1h\text{Mpc}^{-1}}\right)^2
\end{align}
In both cases we see that a co-moving scale of $k/a = \mathcal{O}(0.1h \text{Mpc}^{-1})$ (a linear scale in GR) will experience a very large correction if  $x \ll 1$ is not satisfied.  With $x\ll 1$,  we also have
\begin{align}\label{a_constraint}
A(\varphi) - 1 =  x\left(\frac{\varphi}{\varphi_{\rm min}}\right)^{\frac{1}{\beta}} < x \ll 1
\end{align}
justifying our claim.

To get a constraint on the model parameters we define $k_{\rm mod}$ via
\begin{align}\label{kmoddef}
 \left(\frac{G_{\rm eff}(k_{\rm mod},z=0)-G}{G}\right)_{\rm lin} \equiv 1
\end{align}
and impose $k_{\rm mod} > 0.1h \text{Mpc}^{-1}$ in order for the growth of perturbations to be in agreement with $\Lambda$CDM at large scales. With this definition we can get a constraint on the energy density in the supersymmetron to the total dark energy today. By using Eq.~(\ref{kmoddef}) and Eq.~(\ref{rhomin}) we find
\begin{align}
\frac{\Omega_{\varphi}}{\Omega_{\varphi}+\Omega_{\rm CC}}= \left\{
	\begin{array}{cc}
		 \frac{(n+1)^2\Omega_{\rm CDM}^210^{-4}}{6(1-\Omega_{\rm CDM})}\left(\frac{0.1h/\text{Mpc}}{k_{\rm mod}}\right)^2&  \rho_{\rm CDM}^0 \gg \rho_{\infty}\\
		\frac{n\Omega_{\rm CDM}^210^{-4}(1+z_{\infty})^3}{12(1-\Omega_{\rm CDM})}\left(\frac{0.1h/\text{Mpc}}{k_{\rm mod}}\right)^2& \rho_{\rm CDM}^0 \lesssim \rho_{\infty}
	\end{array}\right.
\end{align}
If $\rho_{\rm CDM}^0 \gg \rho_{\infty}$ the energy density in the supersymmetron today is negligible compared to the CC. In the other regime $ \rho_{\rm CDM}^0 \lesssim \rho_{\infty}$ we find
\begin{align}
\frac{\Omega_{\varphi}}{\Omega_{\varphi}+\Omega_{\rm CC}}  < 10^{-4}(1+z_{\infty})^3
\end{align}
and we must require $z_{\infty} > 10$ if the supersymmetron is to account for a significant part of the dark energy budget. However, the dominating contribution to dark energy must be the CC as otherwise the equation of state Eq.~(\ref{eom_de}) reads $\omega_{\rm DE} \approx 0$ today and hence no acceleration. Thus is both cases we find that a pure CC is required to account for dark energy.
\\\\
Returning to the linear perturbations, we see that the linear effective gravitational constant is increasing as we go to smaller scales (large $k$). For $k > k_{\rm mod}$ the supersymmetron fifth-force is dominating over gravity at the linear level and to have a viable theory we need a large $k_{\rm mod}$. At non-linear scales we expect a chameleon-like effect to kick in and screen the fifth-force. We will study the non-linear effects by looking at the spherical collapse.

%%%%%%%%%%%%%%%%%%%%%%%%%%%%%%%%%%
%% SPHERICAL COLLAPSE
%%%%%%%%%%%%%%%%%%%%%%%%%%%%%%%%%%

\subsection{Spherical collapse}

In this secion we look at the collapse of a spherical top-hat overdensity taking the supersymmetron fifth-force into account. This will allow us to extract constraints on the model parameters by requiring the model to agree with $\Lambda$CDM on large scales.

The equation of motion of a spherical shell at the edge of the top-hat overdensity in a scalar-tensor theory with a fifth-force was derived in \cite{Brax:2010tj}. The final form of the equation can be understood from a simple Newtonian argument. In the derivation below we neglect the CC energy density because the Newtonian picture does not assign any energy density to pressure and therefore a Newtonian derivation cannot yield the correct contribution (which involves contributions from pressure) from the CC without some ad-hoc assumptions.

The total energy of a collapsing spherical shell of matter is given by
\begin{align}
\frac{E}{m_{\rm shell}} = \frac{1}{2}\dot{r}^2 - \frac{GM_{<r}}{r} + V(\varphi)
\end{align}
where $V(\varphi)$ is the fifth-force potential. Neglecting shell-crossing so that  the total energy is conserved and using $\dot{E} = 0$ we get Newton's law for the shell
\begin{align}
m_{\rm shell}\ddot{r} &= -(F_{\rm gravity}(r) + F_{\varphi}(r))
\end{align}
which can be written
\begin{align}\label{newton_sph}
\frac{\ddot{r}(t)}{r(t)} &= -\frac{1}{6}\frac{\rho_{\rm CDM}}{M_{\rm pl}^2}\left(1 + \frac{F_{\varphi}(r)}{F_{\rm gravity}(r)}\right)\nonumber\\
&=-\frac{1}{6}\frac{\rho_{\rm CDM}}{M_{\rm pl}^2}\left(\frac{G_{\rm eff}(r)}{G}\right)_{\rm sph}
\end{align}
The term on the right hand side of Eq.~(\ref{newton_sph}) agrees with the matter term found from a full derivation including pressure and  gives the result
\begin{align}\label{sph_eq}
\frac{\ddot{r}(t)}{r(t)} = \frac{1}{3}\frac{\rho_{\rm DE}}{M_{\rm pl}^2} - \frac{1}{6}\frac{\rho_{\rm CDM}}{M_{\rm pl}^2}\left(\frac{G_{\rm eff}(r)}{G}\right)_{\rm sph}
\end{align}
In the following, the DE density is taken to be a pure CC and the effective gravitational constant is derived in Sec.~(\ref{static}), see Eq.~(\ref{geff_sph}). For a small overdensity of size $r$ we can write Eq.~(\ref{geff_smalldelta}) as
\begin{align}
\left(\frac{G_{\rm eff}(r)}{G}\right)_{\rm sph} &\approx 1 + \left(\frac{4}{n\Omega_{\rm CDM}^0}\right)\times \frac{x 10^{36}}{(1+z_{\infty})^3}\nonumber\\
&\times\left(\frac{\text{Mpc}/h}{r}\right)\times\left(\frac{m_{\infty}}{\text{eV}}\right)\times\left(\frac{\rho_{\infty}}{\rho_{\rm CDM}}\right)^{\frac{2-2\beta}{n+1}}
\end{align}
when $\rho_{\rm CDM} \lesssim \rho_{\infty}$ and
\begin{align}
\left(\frac{G_{\rm eff}(r)}{G}\right)_{\rm sph} &\approx 1 + \left(\frac{4}{\sqrt{2n(n+1)}\Omega_{\rm CDM}^0}\right)\times \frac{x 10^{36}}{(1+z_{\infty})^3}\nonumber\\
&\times\left(\frac{\text{Mpc}/h}{r}\right)\times\left(\frac{m_{\infty}}{\text{eV}}\right)\times\left(\frac{\rho_{\infty}}{\rho_{\rm CDM}}\right)^{\frac{4-2\beta+n}{2(n+1)}}
\end{align}
for $\rho_{\rm CDM} \gg \rho_{\infty}$. Note that the effective gravitational constant in the spherical collapse is much larger than the corresponding linear value. For a large-scale overdensity today, $\text{Mpc}/h \lesssim r$, to agree with $\Lambda$CDM we must require
\begin{align}\label{x_constraint_sph}
x \lesssim 10^{-36}\left(\frac{\text{eV}}{m_{\infty}}\right) \ll 10^{-36}
\end{align}
For such a small value of $x$, by looking at Eq.~(\ref{geff_lin1}-\ref{geff_lin2}) we see that the linear perturbations will be indistinguishable from $\Lambda$CDM. This also means that the adiabatic instability that might exist in these models are avoided at the linear level.

By changing coordinates to $y = \frac{r}{aR}$ where $R = \frac{r_i}{a_i}$ we can write Eq.~(\ref{sph_eq}) in the form
\begin{align}
y'' + &\left(2-\frac{3}{2}\Omega_m(N)\right)y' \nonumber\\&+ \frac{\Omega_m(N)}{2}\left(y^{-3}-1\right)y\left(\frac{G_{\rm eff}(aRy)}{G}\right)_{\rm sph}=0
\end{align}
where a prime denotes a derivative with respect to $N = \log(a)$.
\\\\
The density contrast $\Delta = \frac{\rho_{\rm CDM}}{\overline{\rho}_{\rm CDM}}-1$ of the collapsing sphere can be obtained from $\Delta = y^{-3}-1$ and the mass from $M \approx \frac{4\pi R^3}{3}\overline{\rho}_{\rm CDM}^0$. Early on we have $y\approx 1- \frac{\Delta}{3}$ as the overdensity follows the expansion, and by linearizing this equation we obtain the equation for the linear evolution of the density constrast
\begin{align}
\Delta'' &+ \left(2-\frac{3}{2}\Omega_m(N)\right)\Delta' \nonumber\\&- \frac{3}{2}\Omega_m(N)\Delta \left(\frac{G_{\rm eff}(r)}{G}\right)_{\rm lin}=0
\end{align}
where
\begin{align}
\left(\frac{G_{\rm eff}(r)}{G}\right)_{\rm lin} = 1+2(A_{,\varphi}M_{\rm pl})^2 (1+m_b r)e^{-m_b r}
\end{align}
which is the same equation as for the linear perturbations Eq.~(\ref{linear_pert}) in real space. As mentioned before
\begin{align}
\left(\frac{G_{\rm eff}(r)-G}{G}\right)_{\rm lin} \ll \left(\frac{G_{\rm eff}(r)-G}{G}\right)_{\rm sph} \lesssim \mathcal{O}(1)
\end{align}
on linear scales and the linear equation reduces to that of $\Lambda$CDM. This also shows that non-linear effects are very dominant in this model.

\begin{figure}
\centering
\includegraphics[width=\columnwidth]{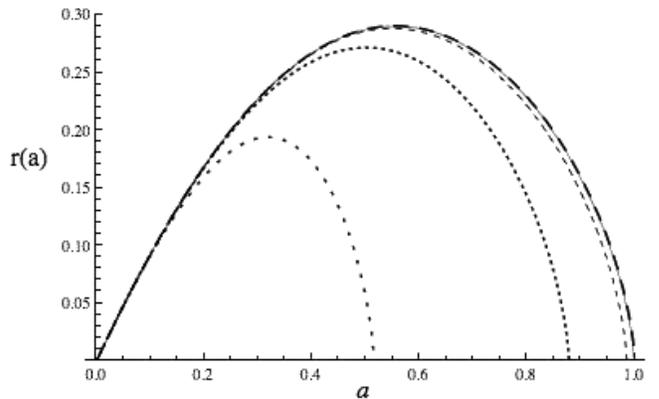}
\caption{$r(a)$ as a function of the scale factor $a$ for $R = \frac{r_i}{a_i} = 0.1,1,10,100$ Mpc$/h$ (from left to right). compared with the behaviour for $r$ in usual $\Lambda$CDM (solid). The initial density contrast is the same in all runs and is fixed such as to give collapse today for $\Lambda$CDM. The supersymmetron parameters are $z_{\infty} = 0.0$, $\beta = n = 1$, $x = 10^{-43}$ and $m_{\infty} = 10^5$eV.}
\label{collapse}
\end{figure}

The initial conditions for the numerical implementation are taken to be the same as for $\Lambda$CDM:
\begin{align}
y_i &= 1 - \frac{\Delta_i}{3},~~~y_i' = \frac{\Delta_i}{3},~~~\Delta_i' = \Delta_i
\end{align}
In Fig.~(\ref{collapse}) we show the evolution of the radius of an overdensity at different scales. Smaller overdensities collapse earlier as the fifth-force is more dominant.

\begin{figure}
\centering
\includegraphics[width=\columnwidth]{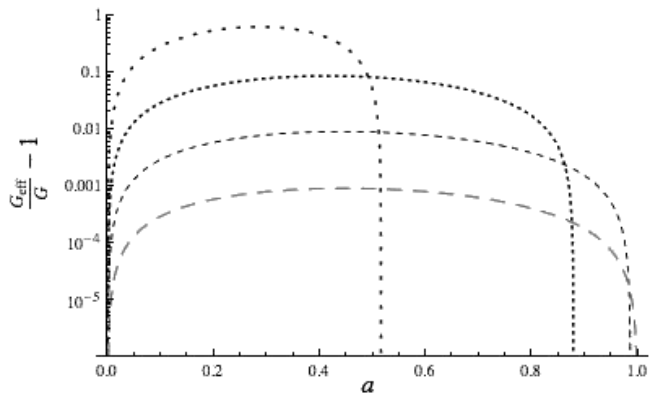}
\caption{The strength of the fifth-force at the surface of the spherical overdensity during the collapse as a function of the scale factor $a$ for $R = \frac{r_i}{a_i} = 0.1,1,10,100$ Mpc$/h$ (from top to bottom). The  parameters are the same as in Fig.~(1).}
\label{geff_fig}
\end{figure}

In Fig.~(\ref{geff_fig}) we plot the evolution of the effective gravitational constant for the same case as Fig.~(\ref{collapse}). As the density contrast of the collapsing sphere increases, the chameleon mechanism kicks in and effectively shields the fifth-force.

\begin{figure}
\centering
\includegraphics[width=\columnwidth]{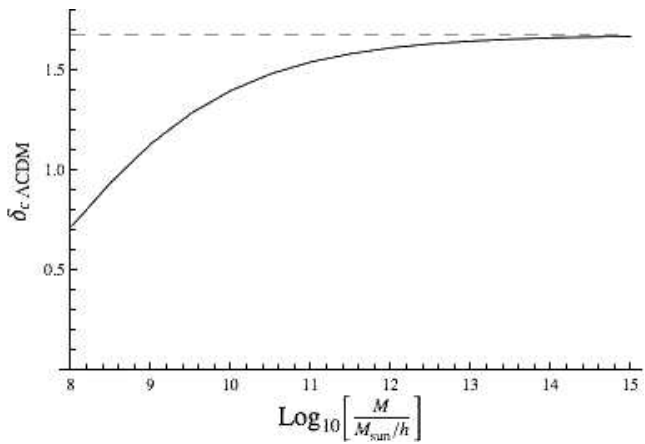}
\caption{The $\Lambda$CDM-lineary-extrapolated critical density contrast for collapse for the supersymmetron as a function of the halo mass. The dashed line shows the $\Lambda$CDM prediction $\delta_c \approx 1.67$. The supersymmetron parameters are $z_{\infty} = 0.0$, $\beta = n = 1$, $x = 10^{-43}$ and $m_{\infty} = 10^5$eV.}
\label{crit_density}
\end{figure}

Too see more clearly the effect of the fifth-force on the formation of halos, we calculate the linearly-extrapolated density contrast for collapse today as function of the virial mass of the halo compared to the $\Lambda$CDM prediction $\delta_c \approx 1.67$, see Fig.~(\ref{crit_density}). Low mass halos are seen to require a smaller linear density contrast than that of $\Lambda$CDM in order to collapse due to the fifth-force.

With the linear collapse threshold $\delta_c$, we can predict the
halo mass function. In the standard Press-Schechter approach one assumes
that all regions with $\delta > \delta_c$ in the linear extrapolated
density field collapse to form halos. The fraction of mass within halos
with a given mass is determined by the variance of the linear density
field smoothed over that scale. We adopt the Sheth-Tormen (ST)
prescription \cite{Sheth:1999mn} for the halo mass function. The ST
description for the co-moving number density of halos per logarithmic
mass interval in the virial mass $M$ is given by
\begin{align}
n_{\log M} = \frac{dn}{d\log M} = \frac{\overline{\rho}}{M} f(\nu)
\frac{d\nu}{d \log M}
\end{align}
where the peak threshold
\begin{align}
\nu = \frac{\delta_c(M)}{\sigma(M)}
\end{align}
and
\begin{align}
\nu f(\nu) = C \sqrt{\frac{2}{\pi}a\nu^2} \left(1 +
\frac{1}{(a\nu^2)^p}\right) e^{-\frac{a\nu^2}{2}}
\end{align}
We adopt the standard parameters $a=0.75$ and $p=0.3$ in the following
for which $C=0.322$.
$\sigma(M)$ is the variance of the linear density field convolved with a
top-hat of radius $R$ ($M = \frac{4\pi}{3}\overline{\rho}R^3$)
\begin{align}
\sigma(R)^2 = \int \frac{k^3}{2\pi^2}P_L(k) |W(kR)|^2 \frac{dk}{k}
\end{align}
where $P_L(k)$ is the linear power-spectrum, which for the
supersymmetron is that of $\Lambda$CDM, and $W(x) = \frac{3}{x^3}\left(\sin(x) - x\cos(x)\right)$ is the Fourier transform
of the top-hat window function. The power-spectrum is normalized to
$\sigma_8 \equiv \sigma(R=8\text{Mpc}/h)$. We have chosen $\sigma_8 =
0.8$ in our analysis.

The virial theorem gives us the condition for virialization of a halo.
This condition reads $2T + W = 0$, where $T=\frac{3}{10}M\dot{R}^2$ is
the kinetic energy and $W = - \int d^3x\rho_m(x) \vec{x}\cdot
\vec{\nabla}\Psi$ is the potential energy. In the presence of a
fifth-force the potential $\Psi = \Phi_N + A(\varphi \rho_m)$ is the sum of the
gravitational potential and the fifth-force potential. In the spherical
symmetric case,  we have
\begin{align}
W = W_N - \int_0^{R} 4\pi r^3 \rho_m \frac{d A(\varphi)}{dr}dr
\end{align}
where $W_N = -\frac{3}{5}\frac{GM^2}{R}$ is the gravitational potential
energy. Performing the integration on the r.h.s. using integration by
parts we find
\begin{align}
\left|\int_0^{R} 4\pi r^3 \rho_m \frac{d A(\varphi)}{dr}\right| \leq 4\pi
R^3\rho_m(A(\varphi( R ))-1)
\end{align}
Note that this term is usually much smaller than the gravitational
potential energy. This can be understood from the chameleon thin-shell
analogue: the fifth-force is only felt in a thin-shell close to the
surface of the body and therefore the potential energy associated with
the fifth-force for the whole halo is small.

\begin{figure}
\centering
\includegraphics[width=\columnwidth]{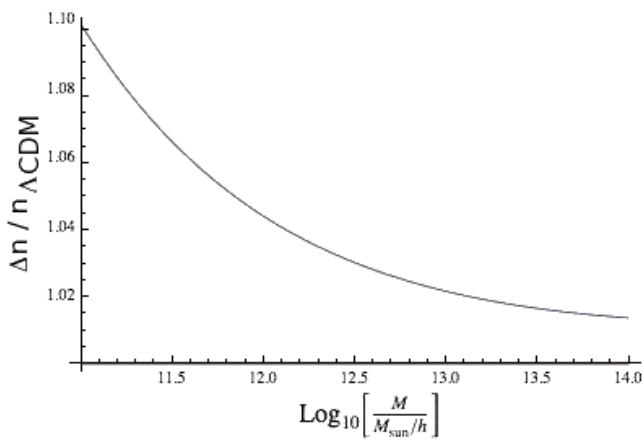}
\caption{The fractional difference in the supersymmetron mass function
compared to $\Lambda$CDM at $z=0$. The symmetron parameters are
$m_{\infty} = 10^5$eV, $z_{\infty} = 0.0$, $x=10^{-43}$ and $\beta=n=1$.
The supersymmetron converges to  $\Lambda$CDM for large halo masses.}
\label{mass_function}
\end{figure}

In Fig.~(\ref{mass_function}) we show the ST mass function of the
supersymmetron relative to that of $\Lambda$CDM. Because the
supersymmetron fifth-force is increasing with decreasing scale we
recover $\Lambda$CDM on large scales, but see an enhancement in the
mass-function for low mass halos.

%%%%%%%%%%%%%%%%%%%%%%%%%%%%%%%%%%
%%%%%%%%%%%%%%%%%%%%%%%%%%%%%%%%%%
%%%%%%%%%%%%%%%%%%%%%%%%%%%%%%%%%%
%% THE ORIGINAL MASS SCALES
%%%%%%%%%%%%%%%%%%%%%%%%%%%%%%%%%%
%%%%%%%%%%%%%%%%%%%%%%%%%%%%%%%%%%
%%%%%%%%%%%%%%%%%%%%%%%%%%%%%%%%%%

\section{ Mass scales}\label{sec4}

Having found a range for our model parameters which gives predictions that are in agreement with current observations, we will now analyse how these constraints  affect the  original mass-scales $\Lambda_i$ of the model.
\\\\
We start by defining $x' = 10^{40}x$, $m' = \frac{m}{g\text{TeV}}$ and $z' = (1+z_{\infty})^3$. We can now rewrite Eq.~(\ref{Xeq}) as
\begin{align}
\frac{\phi_{\rm min}}{H_0} \approx  10^{5}m' x'
\end{align}
and Eq.~(\ref{Lambdaeq}) as
\begin{align}
\left(\frac{\Lambda}{M_{\rm pl}}\right)^4 \approx 10^{-160}x'z'
\end{align}
From Eq.~(\ref{phimineq}) we find
\begin{align}
\left(\frac{M}{M_{\rm pl}}\right)^{n+4} \approx 10^{-160-55n}x'^{n+1}m'^nz'
\end{align}
By using  Eq.~(\ref{scaleseq}) we get
\begin{align}\label{scales}
\left(\frac{\Lambda_1}{\Lambda_0}\right)^{2\beta-2}\left(\frac{\Lambda_0}{M_{\rm pl}}\right)^{n+4} &\approx 10^{-160-55n}x'^{n+1}m'^nz'\\
\left(\frac{\Lambda_1}{\Lambda_2}\right)^{2\beta-2}\left(\frac{\Lambda_2}{M_{\rm pl}}\right)^{4} &\approx 10^{-160}x'z'
\end{align}
from which we find
\begin{align}
\left(\frac{\Lambda_2}{\Lambda_0}\right)^{2 + 2\beta}\left(\frac{\Lambda_0}{10^5 H_0}\frac{1}{x'm'}\right)^{n} \approx 1
\end{align}

The simplest case to analyze is $\beta = 1$ for which the scale $\Lambda_1$ vanishes from the theory. We find
\begin{align}
\Lambda_2 \approx 10^{20}(x'z')^{1/4} H_0
\end{align}
i.e. $\Lambda_2$ needs to be between the current Hubble scale and the dark energy scale. This scale can be elevated by increasing the redshift $z_{\infty}\gg 1$, but we typically need a redshift in the very early Universe to reach super-TeV scales. For $\Lambda_0$ we find
\begin{align}
&\Lambda_0 \approx 10^{20}\left(\frac{z'}{(x'm')^n}\right)^{\frac{1}{4}}  H_0
\end{align}
In the general case $\beta \not= 1$ we see from Eq.~(\ref{scales}) that taking $\beta < 1$ together with $\Lambda_1 \gg \Lambda_0,\Lambda_2$ can serve to increase the other two scales. For example $\beta = \frac{1}{2}$ and $\Lambda_1 \approx M_{\rm pl}$ gives
\begin{align}
&\Lambda_2 \approx \Lambda_0 \approx 10^{28}(x'z')^{\frac{1}{5}}  H_0
\end{align}
which is around the dark energy scale. There seems to be no unfine-tuned way  of bringing these mass-scales up to typical particle physics scales if we want the cosmological symmetry breaking to be close to the present era. For example to have $\Lambda_2 \sim$ TeV when $\beta=1$ and $x'\approx 1$ then Eq.~(\ref{scales}) shows that we need $z' \approx 10^{100}$ which translates to $\rho_{\infty} \approx (10^{13}\text{GeV})^4$.

Finally and from a field theoretic point of view, the scale $\Lambda_1$ has a different status from $\Lambda_{0,2}$. The former appears in the K\"ahler potential as a suppression of scales for higher dimensional operators and signals the typical scales above which the effective field theory description breaks down. On the contrary, $\Lambda_{0,2}$ appear in the superpotential and are protected by non-renormalisation theorems. Hence we expect that $\Lambda_1$ should be sensitive to high energy physics and represents the effective cut-off of the theory. On the other hand, $\Lambda_{0,2}$ may be already present at very high energy even if these scales are very low. Of course, this does not provide an explanation for the discrepancy of scales between $\Lambda_1$ and $\Lambda_{0,2}$ which is not natural.

%%%%%%%%%%%%%%%%%%%%%%%%%%%%%%%%%%
%%%%%%%%%%%%%%%%%%%%%%%%%%%%%%%%%%
%%%%%%%%%%%%%%%%%%%%%%%%%%%%%%%%%%
%% DISCUSSION AND CONCLUSIONS
%%%%%%%%%%%%%%%%%%%%%%%%%%%%%%%%%%
%%%%%%%%%%%%%%%%%%%%%%%%%%%%%%%%%%
%%%%%%%%%%%%%%%%%%%%%%%%%%%%%%%%%%

\section{Discussion and conclusion}\label{sec5}

We have studied the cosmological evolution of the supersymmetron and its possible effects on structure formation. Requiring that  linear perturbations are in agreement with $\Lambda$CDM on  large scales, we find that the energy density in the supersymmetron is negligible compared to the dark matter density and a pure cosmological constant must be introduced to play  the role of  dark energy.

The non-linear evolution of the model was also investigated by using the spherical collapse model. Spherically symmetric solutions to the field equation have been  derived and used to predict the fifth-force effects on a collapsing halo. The effective gravitational constant at the edge of a spherical overdensity has been  found to be much larger than the linear prediction due to the highly non-linear properties of the model. The model parameter must be tuned such that the spherical collapse is under control on large scales. This implies that linear perturbations reduce to that of $\Lambda$CDM. On non-linear scales the model then predicts a faster collapse than that of $\Lambda$CDM. In particular we find that the supersymmetron predicts an excess of small mass halos compared to $\Lambda$CDM. However for this to be the case, one or more of the mass-scales in the theory must be fine-tuned.

On very small scales, i.e. galaxy-scales, the matter density is large enough to effectively screen the fifth-force via the chameleon mechanism.
This non-linear regime could in principle be probed using N-body simulations. However, due to the enormous mass of the field this poses a severe challenge for existing methods.

%%%%%%%%%%%%%%%%%%%%%%%%%%%%%%%%%%
%%%%%%%%%%%%%%%%%%%%%%%%%%%%%%%%%%
%%%%%%%%%%%%%%%%%%%%%%%%%%%%%%%%%%
%% ACKNOWLEDGEMENT
%%%%%%%%%%%%%%%%%%%%%%%%%%%%%%%%%%
%%%%%%%%%%%%%%%%%%%%%%%%%%%%%%%%%%
%%%%%%%%%%%%%%%%%%%%%%%%%%%%%%%%%%

\section{Acknowledgement}
A.C.D. is supported in part by STFC. H.A.W. thanks the Research Council of Norway FRINAT grant 197251/V30. H.A.W. thanks DAMPT at Cambridge University and IPhT CEA Saclay for the hospitality where a part of this work was carried out.

%%%%%%%%%%%%%%%%%%%%%%%%%%%%%%%%%%
%%%%%%%%%%%%%%%%%%%%%%%%%%%%%%%%%%
%%%%%%%%%%%%%%%%%%%%%%%%%%%%%%%%%%
%% REFERENCES
%%%%%%%%%%%%%%%%%%%%%%%%%%%%%%%%%%
%%%%%%%%%%%%%%%%%%%%%%%%%%%%%%%%%%
%%%%%%%%%%%%%%%%%%%%%%%%%%%%%%%%%%

\end{document}